\documentstyle[epsf,graphicx]{mn}

\def\etal{{et al.\ }}

\def\x2{$\chi^{2}$}

\def\asca{{\it ASCA }}
\def\rosat{{\it ROSAT }}

\def\chandra{{\it Chandra }}

\def\bepposax{{\it BeppoSAX }}
\def\xmm{{\it XMM-NEWTON }}
\def\gis{{\it GIS }}

\def\x2{$\chi^{2}$}

\def\funits{$\rm{erg\,s^{-1}\,cm^{-2}}$}
\def\cunits{$\rm{cm^{-2}}$}

\newbox\grsign \setbox\grsign=\hbox{$>$} \newdimen\grdimen \grdimen=\ht\grsign
\newbox\simlessbox \newbox\simgreatbox \newbox\simpropbox
\setbox\simgreatbox=\hbox{\raise.5ex\hbox{$>$}\llap
     {\lower.5ex\hbox{$\sim$}}}\ht1=\grdimen\dp1=0pt
\setbox\simlessbox=\hbox{\raise.5ex\hbox{$<$}\llap
     {\lower.5ex\hbox{$\sim$}}}\ht2=\grdimen\dp2=0pt
\setbox\simpropbox=\hbox{\raise.5ex\hbox{$\propto$}\llap
     {\lower.5ex\hbox{$\sim$}}}\ht2=\grdimen\dp2=0pt
\def\simgreat{\mathrel{\copy\simgreatbox}}
\def\simless{\mathrel{\copy\simlessbox}}
\def\simprop{\mathrel{\copy\simpropbox}}

\begin{document}

\title[ \asca observations of deep \rosat fields V] {\asca observations of
deep \rosat fields V. The X--ray spectrum of hard X-ray selected QSOs.}

\author[A. Pappa et al.]
       {A. Pappa$^{1,}\thanks{Present address: Institute for Astronomy,
University of Edinburgh, Royal Observatory, Blakford Hill, Edinburgh
EH9 3HJ.} $, G.C. Stewart$^{1}$, I. Georgantopoulos$^{2}$,
 R.E. Griffiths$^{3}$, B.J. Boyle$^{4}$ and \\ \LARGE{T. Shanks$^{5}$}  \\  
 Department of Physics and Astronomy, University of Leicester, 
Leicester, LE1 7RH \\
 National Observatory of Athens, Lofos Koufou, Palaia Penteli, 
15236, Athens, Greece  \\
 Physics Department, Carnegie Mellon University, Pittsburgh, PA \\
 Physics department, University of Durham, Science Labs, South Road, DH1 3LE\\
 Anglo-Australian Observatory, Epping, Australia 
} 

\maketitle

\label{firstpage}

\begin{abstract}
We present an analysis   of the \rosat and \asca spectra of  21 broad
line  AGN  (QSOs) with $z\sim 1$ detected in the 2-10 keV band 
 with the \asca \gis.
The summed spectrum in the \asca band is well described by a power-law with
$\Gamma=1.56\pm0.18$, flatter that the average spectral index
 of  bright QSOs and consistent with the spectrum 
 of the X-ray background in this band. 
The flat spectrum in the \asca band could be explained by only a moderate absorption  
($\sim 10^{22} \rm cm^{-2}$) assuming the typical AGN spectrum 
 ie a power-law with $\Gamma$=1.9. 
This could in principle 
 suggest that some of the highly obscured AGN, required by most 
 X-ray background synthesis models, may be associated 
 with  normal blue QSOs rather than  narrow-line AGN.
  However, 
 the combined 0.5-8 keV \asca-\rosat spectrum is well fit by a power-law of 
$\Gamma=1.7\pm0.2$ with a spectral upturn at soft energies.
 It has been pointed out that such an upturn 
 may be an artefact of uncertainties in the calibration 
 of the ROSAT or ASCA detectors. Nevertheless  if real, it could imply
that the above absorption model suggested by the \asca data alone is
ruled out. Then 
 a large fraction of QSOs could have ``concave'' spectra ie they 
 gradually steepen towards softer energies. 
This result is in agreement with the \bepposax hardness ratio analysis
 of $\sim$ 100 hard X-ray selected sources.

\end{abstract}

\def\simgreat{\mathrel{\copy\simgreatbox}}
\def\simless{\mathrel{\copy\simlessbox}}
\def\simprop{\mathrel{\copy\simpropbox}}


\section{INTRODUCTION}
It has been almost 40 years since Giacconi \etal (1962) 
discovered the X-ray Background (XRB), 
the first cosmic background detected.  Nevertheless  its origin is
still  not fully understood.
At a flux limit of $\sim 10^{-15} \rm erg \,cm^{-2}  s^{-1}\, 
70-80\% $ of the soft 0.5-2 keV XRB is resolved into 
discrete sources and the majority of these  are QSOs ie broad 
line AGN (Schmidt \etal 1998).
 However, the spectrum of nearby, bright  AGN
 is inconsistent with the spectrum of the 
 X-ray background.
 This is known as the ``spectral paradox'' (Boldt 1987).  
 Indeed,  {\it HEAO-1}, \asca and \bepposax 
 observations have shown that the XRB spectrum in the 1 - 10 keV band 
 has a spectral index of $\Gamma=1.4$ (Marshall et al. 1980, 
 Gendreau et al. 1995, Miyaji et al. 1998, Vecchi et al. 1999). 
 On the other hand the nearby radio-quiet QSOs
 typically have a significantly softer spectrum, with  $\Gamma\sim1.8$,
 (Lawson et al. 1997, Reeves \etal 1997)
 and therefore they can contribute only a small fraction 
 of the XRB. 

In  the hard X-ray band (2-10 keV) only  $\sim 30\%$ of  the 
XRB has been resolved into discrete 
sources (with \asca and \bepposax) 
 down to a flux limit of $\sim 5\rm x10^{-14} \rm erg \, 
 cm^{-2}s^{-1} $ (Georgantopoulos \etal 1997, Cagnoni et al. 1998,
 Giommi et al. 2000, Akiyama  \etal 2000). 
 The number-count distribution, logN-logS in the
2-10 keV band, is 
about a factor of two above the \rosat
counts, assuming a photon spectral index of $\Gamma=2$ for the \rosat sources. 
This suggests the presence of a population with flat or
absorbed spectra.
Setti $\&$ Woltjer (1979) first suggested that the entire 
extragalactic X-ray  light could be
explained in the context of a simple unified scheme for AGNs. Several
models have been 
developed 
 along these lines (e.g. Comastri \etal 1995)
 which provide good fits to the XRB and the X-ray source counts.
All these  assume that the XRB consists of a
mix of AGNs obscured by a range of column densities.
 However, only a small number of obscured high redshift 
 AGN have been detected so far in \asca and \bepposax surveys
 (eg Ohta et al. 1996, Boyle et al. 1998a, Georgantopoulos et al. 1999,
 Fiore et al. 1999). 

The deep {\it Chandra} surveys deepened the riddle of the 
 origin of the XRB even further. 
 In the hard 2-10 keV band they probed fluxes at least an order 
 of magnitude deeper than  \asca (Mushotzky et al. 2000)
 albeit with limited number statistics due to the small 
 field-of-view of ACIS onboard {\it Chandra}. 
 A large fraction of the detected sources is associated with 
 QSOs which appear to have steep spectra. 
 Surprisingly, no numerous, clearcut examples of the 
 putative obscured AGN population at high
 redshift have yet been found. 
 Instead, two 'new' populations emerged which are  associated with 
 either early-type galaxies or extremely faint optical counterparts. 

Here, we derive the broad-band (0.5-8 keV) spectral properties of the 
'typical' (ie high redshift, faint) hard X-ray selected 
QSOs in our \asca survey. These contribute 
a large fraction  of the XRB and therefore comparison 
of their spectrum with that of the XRB is expected to shed 
more light on the spectral paradox.   
We note that our objects span a wide range of redshifts,
and hence no physical significance should be attributed to the models
applied and no straightforward constraints on the AGN accretion 
physics can be derived. 
 In contrast the aim of our analysis is to parameterise the 
 ``average'' QSO spectrum over a large redshift range and 
 compare it with that of the cosmic XRB.  

\section{THE SAMPLE}
We have performed an \asca follow-up (Georgantopoulos et al. 1997) 
of our deep \rosat  survey 
(Georgantopoulos et al. 1996).
We have observed 6 fields (SGP2, QSF1, QSF3, GSGP4,
BJS855, BJS864) and in our first, quick-look analysis, 
  we detected 39 sources down 
to a flux limit of $ \rm S_{(2-10 keV)}\rm \sim 5x10^{-14}erg \,
cm^{-2}s^{-1}$ in the 2-10 keV band. 
We obtained  optical identifications 
for the vast majority of our sources using the 3.9m AAT telescope. 
Most (21 objects) are  QSOs. All but one (AXJ1343.5-0004) have been detected in the
soft X-rays by \rosat. AXJ1343.5-0004 is identified as a UV excess
quasar from Boyle \etal 1990.
 Their redshifts range from z=0.145 to z=1.952
with a mean of z$\sim1$ 
(see Georgantopoulos \etal 1997 and Boyle et al. 1998b)
There is also evidence for the  
presence of QSOs obscured in X-rays with some modest  
amount of reddening. These objects have  narrow  
lines in the optical range but they clearly present broad lines in
their infrared spectra (eg Georgantopoulos et al. 1999). 
 
Only a small fraction of our sources are radio-loud.
 We cross correlated the list of our QSOs with the NRAO
VLA Sky Survey (NVSS) list of radio sources (Condon \etal 1998). The NVSS covers the sky
north of J2000 $\delta=-40^o$ at 1.4 GHz down to a flux limit of
S=2.5 mJy. Therefore it does not cover our QSF1 and QSF3 fields. 
Of 12 QSOs in the other three fields, one has a possible NVSS 
counterpart within 1 arcmin. The probability of finding a source
within 1 arcmin of an arbitrary position is 0.0145.
Boyle \etal (1993, 1995) made a deep observation at 1.472 GHz with the
Australia Telescope Compact Array of the QSF3 field.
They found 6 coincidences within 15 arcsec. 
Out of our 6 QSOs in the field only one (AX J0342.0-4403) is associated with a
 radio source.
Hereafter, we include the two (possible) radio-loud QSOs in our analysis.

\section{Data reduction and analysis}

We observed 6 fields from our \rosat survey with \asca  (Tanaka, Inoue
$\&$Holt 1994). Here we
present the analysis of the GIS data alone, because (a) the GIS field
of view matches that used in our \rosat survey and (b) with the GIS we
maximize the effective exposure times after rejecting time periods
with high rates of particle events.
In table 1. we give the field names in column (1), equatorial
coordinates (J2000) in columns (2) and (3), the Galactic hydrogen column
density in units of $10^{20}$\cunits (Stark \etal 1992) in column
(4), the number of QSOs detected in each field (5), 
the net \asca exposure times per field in ks in column (6) and the net
\rosat exposure times in ks in column (7).
The QSF3 field was observed with \asca in 1993 June 
and 1993 September during the
period of performance verification phase (PV phase). The GSGP4,
BJS855, QSF1, SGP2 and BJS864 fields
were observed in 1994 June, 1995 November, 1997 January, 1997 July and
January 1998 respectively. Details of the \rosat observations 
are given in Blair \etal 2000. 
In table 2 we present the list of QSOs. The names are given in column (1), their \asca
position in column (2), the redshift of their optical counterpart (3),
the count rate in the 1-2 keV and 2-10 keV bands in (4) and (5)
respectively and their 2-10 keV flux in column (6); in column (7) the
hardness ratio; 
in column (8)
the radio flux (mJy) is given. The X-ray fluxes are 
estimated for a single power-law model with $\Gamma$=1.56 and Galactic
absorption. This model was chosen because it describes the \asca data
adequately (see $\S$4.2). If, on the other hand, we use the standard 
 AGN spectrum, namely a single power-law with
$\Gamma$=1.7, the obtained fluxes are $\sim5\%$ lower. 

We used the standard 'Revision 2' ( see the \asca Data Reduction
Guide) processed data from the Goddard
Space Flight Center (GSFC). 
 As our sources are quite faint, 
we used a circular source region centered
on the source of only 1 arcmin radius, which includes 33 per cent
of the total energy.
In addition, by using such a small region we avoid overlapping of
extraction regions, since some of our sources lie close to each other.
Background counts were estimated
from a source-free circular region centered in the field
of view of the GIS detector. Because our sources lie at
different off-axis angles we took into account the vignetting, by
correcting the source counts for this effect. We also corrected the counts
 taking into account the light falling out of the 1 arcmin extraction radius.
The spectral fitting was performed with XSPEC v10.
The spectra were rebinned such that each resultant channel had at
least 20 counts per bin (source+background), which permitted us to use $\chi^2$
minimizations for spectral fitting.
Due to considerable contamination from the Galactic background
as well as some uncertainties in the
low energy calibration of \asca ( George \etal 1998) we restricted our
spectral analysis to the 0.8-8.0 keV energy band. Above 8 keV the
signal-to-noise drops rapidly and thus we choose to ignore these data.

\begin{table*}
\caption{List of \asca fields} 
\begin{tabular}{lcccccc} 

Field        & $R.A.$       & $Dec.$   & $N_H$ & number &\asca exposure & \rosat exposure\\ 
             &J2000             &J2000         &($\times
10^{20}$\cunits)& of QSOS &(ksec) & (ksec)  \\ 
(1)&(2)&(3)&(4)&(5)&(6)&(7)\\\hline
QSF3   &03 41 44.4 & -44 07 04.8 &1.7& 6& 78& 52\\
QSF1   &03 42 10.4 & -44 54 38.5 & 1.7&3 & 44& 49 \\
SGP2   &00 52 08.7 & -29 05 00.6 & 1.8& 3 & 50& 24 \\
GSGP4  &00 57 29.78 & -27 37 21.0 & 1.8& 4& 41& 47\\
BJS855 &10 46 21.36 &  -00 20 17.8 & 1.8& 2& 46& 27 \\
BJS864 &10 46 21.36 &-00 20 17.8   & 1.9& 3& 45& 23\\ \hline\

\end{tabular}
\end{table*}

\begin{table*}
\caption{The list of the QSOs in the 6 fields. The columns contain the
following information: (1) The source name; (2) \asca position of the object;
(3) The redshift of the optical counterpart; (4) \asca GIS count rate
in the 1-2 keV band, together with the photon errors in units of 
$10^{-3}$ count  s$^{-1}$; (5) same as (4) but in 2-10 keV band; (6) \asca GIS flux in the 2-10 keV
band for a power-law with
$\Gamma$=1.56 and Galactic absorption in units of $\times10^{-14}$
\funits; (7) hardness ratio; (8) flux at 1.4 GHz in $m$Jy.}
\begin{tabular}{|c|c|c|c|c|c|c|c|}  

Name  & \asca position &z     & count rate  & count rate
& \asca flux & HR&$S_{1.4 GHz}$   \\ 

   &    &  & 1-2 keV & 2-10 keV & 2-10 keV & &\\ 
(1)& (2) &(3) &(4) &(5) &(6)  &(7)& (8) \\ \hline
AX J0050.8-2902&00 50 53.8\, -29 02 16 & 0.428 &2.1 $\pm$ 0.2 &2.7 $ \pm 0.3 $ & 16.5 $ \pm 1.7 $&+0.1$\pm$0.1&$<2.5$\\   
AX J0051.9-2913&00 51 56.7\, -29 13 56 & 2.056 &0.6$\pm$ 1.3   &1.4 $ \pm 0.2 $ & 8.2  $ \pm 1.1 $&+0.3$\pm$0.3&$<2.5$ \\    
AX J0053.0-2927&00 53 05.1\, -29 11 39 & 0.830 &1.3$\pm$ 0.2   &2.1 $ \pm 0.3 $ & 13.0 $ \pm 1.6$ &+0.3$\pm$0.1&$<2.5$\\   
AX J0056.4-2748&00 56 25.6\, -27 48 48 & 0.145 &9.1$\pm$ 0.7   &6.1 $ \pm 0.5 $ & 30.7 $ \pm 2.8 $&-0.2$\pm$0.1&$<2.5$ \\  
AX J0056.5-2729&00 56 31.1\, -27 29 47 & 1.010 &3.0$\pm$ 0.5   &2.5 $ \pm 0.3 $ & 15.2 $ \pm 1.8 $&-0.1$\pm$0.1&0.086 \\   
AX J0057.3-2731&00 57 20.8\, -27 31 53 & 1.209 &2.8$\pm$ 0.5   &1.9 $ \pm 0.3 $ & 11.7 $ \pm 1.6 $&-0.2$\pm$0.1&$<2.5$ \\  
AX J0057.8-2735&00 57 48.4\, -27 35 56 & 0.57  &$<$0.2         &0.8 $ \pm 0.3 $ & 3.70 $ \pm 1.7 $&$>0.5$&$<2.5$\\ 
AX J0342.4-4511&03 42 27.0\, -45 11 58 & 0.443 &1.7$\pm$ 0.3  &3.0 $ \pm 0.3 $ & 18.1 $ \pm 1.8 $&+0.3$\pm$0.2&- \\   
AX J0343.2-4451&03 43 16.3\, -44 51 44 & 1.410 &1.3$\pm$ 0.3  &2.2 $ \pm 0.2 $ & 13.3 $ \pm 1.5 $&+0.3$\pm$0.1&-\\   
AX J0342.5-4502&03 42 35.0\, -45 02 19 & 0.185 &1.1$\pm$ 0.3  &1.6 $ \pm 0.2 $ & 9.5  $ \pm 1.2 $&+0.2$\pm$0.1&-\\ 
AX J0341.1-4412&03 41 04.5\, -44 12 04 & 1.808 &0.8$\pm$ 0.1  &1.2 $ \pm 0.4 $ & 7.2  $ \pm 2.2 $&+0.2$\pm$0.2&$<0.125^{\star}$\\   
AX J0341.4-4410&03 41 23.0\, -44 10 47 & 0.505 &$<$0.2         &1.0 $ \pm 0.3 $ & 6.1  $ \pm 1.7 $&$>$0.6&$<0.125^{\star}$\\   
AX J0342.0-4410&03 42 01.1\, -44 10 53 & 1.840 &$<$0.1         &1.3 $ \pm 0.3 $ & 7.6  $ \pm 1.7 $&$>$0.8&$<0.125^{\star}$\\   
AX J0342.0-4403&03 42 02.4\, -44 03 51 & 0.635 &$<$0.1         &1.0 $ \pm 0.3 $ & 6.0  $ \pm 1.8 $&$>$0.7&0.142$^{\star}$\\   
AX J0342.3-4412&03 42 19.1\, -44 12 38 & 1.091 &$<$0.3         &0.8 $ \pm 0.3 $ & 4.9  $ \pm 1.9 $&$>$0.2&$<0.125^{\star}$\\   
AX J0342.6-4404&03 42 35.4\, -44 04 41 & 0.377 &0.2$\pm$0.3    &1.4 $ \pm 0.4 $ & 8.4  $ \pm 2.5 $&+0.7$\pm$0.3&$<0.125^{\star}$\\ 
AX J1046.1-0020&10 46 05.1\, -00 20 48 & 1.070 &0.5$\pm$ 0.1  &0.9 $ \pm 0.1 $ & 5.4  $ \pm 0.6 $&+0.3$\pm$0.1&$<2.5$\\   
AX J1046.2-0022&10 46 13.4\, -00 22 16 & 1.952 &$<$0.4         &1.9 $ \pm 0.1 $ & 5.3  $ \pm 0.6 $&$>$0.6&$<2.5$\\   
AX J1344.6-0015&13 44 57.8\, -00 15 09 & 0.244 &6.4$\pm$0.5    &5.9 $ \pm 0.5 $ & 35.6 $ \pm 2.8 $&-0.1$\pm$0.1&$<2.5$\\
AX J1343.5-0004$^{\ast}$&13 43 26.0\, -00 16 14 & 1.511 &$<$0.2         &1.2 $\pm 0.2 $ & 7.0 $ \pm 1.1$&$>$0.6&$<2.5$ \\
AX J1343.3-0016&13 43 51.8\, -00 04 41 & 1.14  &$<$0.3         &2.4 $ \pm 0.3 $ & 14.4 $ \pm 1.7$& $>$0.7&$<2.5$\\ \hline

$^{\star}$ From Boyle \etal 1995. \\
$^{\ast}$ object not detected with \rosat.
\end{tabular}

\end{table*}

\section{X-ray spectrum}

\subsection{Individual source spectra}  

Since the photons from the individual QSOs are too few to give  
reliable spectra, clues for the 
properties of each QSO come from their hardness ratios (HR). 
Here we define the hardness ratio as h-s/h+s, 
where h and s are the total number of counts, 
in the 2-10 keV and 1-2 keV bands respectively.
 In the case where there was no detection 
 in the 1-2 keV band, we estimate the 
 3$\sigma$ upper limit following Kraft et al. (1991).   
We test for possible systematic biases that may arise due to the
combined energy and radial dependence of the PSF, by splitting the entire
sample into sources lying within the 12-arcmin radius from the center
of the GIS and those lying beyond. Additionally, we created simulated
spectra of sources lying at various distances from the center and
estimated their hardness ratio. No trend in HR with off-axis angle is
apparent at any flux and spectral shape.
In fig. 1 the hardness ratio of each  object
versus the observed flux in the 2-10 keV
band is shown. 
The hardness ratios for four different power-law models assuming
Galactic absorption are shown (left hand scale). 
The right hand scale indicates the expected spectrum in the case of 
 $\Gamma=1.9$ for different absorbing column densities. 
One interesting result is that although our objects
are optically classified as  QSOs,
the data require a moderate absorption  
($\sim 10^{22} \rm cm^{-2}$) in order to reproduce the flat 
spectra observed. The number of sources with flat
spectra increases towards faint fluxes, (and hence the mean spectrum flattens)
 suggesting the emergence of a hard X-ray QSO population. 
 This is in agreement with  
 previous results by Ueda \etal (1999) as well as results from hard X-ray selected type I AGN
with \bepposax and \chandra (Della Ceca \etal (1999); Fiore \etal
2001). 
In addition our result is in agreement with the hardness
ratio analysis of $\sim$ 100 hard X-ray selected sources by Giommi \etal 2000.
 
Before interpreting this result we examined whether the observed
flattening is due to systematic effects. 
As our sources are faint a possible source of systematic errors  
is the background subtraction.
Since the spectrum of the X-ray background is flat ($\Gamma\sim$ 1.4), 
any over-subtraction of the background will produce a flat spectrum. 
This effect  will be apparent at faint sources close to the 
limit of the survey, 
since at this faint limit, the sources have fluxes comparable to that
of the  background. 
At higher fluxes this effect will be negligible.
Another source of systematic effects, suggested by Della Ceca \etal 1999, may 
be due to spectral bias in the source selection.
They showed that for sources with same flux but different spectra,
different number of counts will be detected. As we approach 
the flux limit of the survey, sources with a favourable spectrum will
be detected, whereas sources with the same flux but with 
an unfavourable spectrum will be missed. However they showed that in 
the case of the \asca GIS, this selection effect favours the 
detection of steep spectrum sources. 
In addition to this, it is clear from Fig. 1 that the hardening of
the spectra is observed in the brighter data as well.
The last two arguments give support to the reality of the observed trend.

The flattening could be attributed to intrinsic 
 absorption. Indeed, a relatively small amount of 
 absorption ($<10^{22}$\cunits) could easily yield 
 an effective index of   $\Gamma\sim 1.5$. 
 Some high redshift QSOs with high column densities 
 have been found in the HELLAS 5-10 keV \bepposax survey 
(Fiore et al. 2000) and elsewhere (Georgantopoulos et al. 
 1999, Boyle et al. 1998, Halpern et al. 1998,   
 Akiyama et al. 2000, Reeves $\&$ Turner 2000). 
Recently, Norman \etal 2001, detected a further example of a type II
QSO in the \chandra Deep Field South; CDF-S 202.
However the \chandra surveys have failed to detect a large number of clear-cut examples of
this population of object as yet.    
As we have redshifts for   
 our sample we can explore further the origin 
 of this spectral flattening. 
 For example,  Vikhlinin \etal (1995) suggested that 
 absorbed QSO spectra may originate due to 
damped $Ly\alpha$ clouds associated with  protogalaxies. 
 We plot the
 hardness ratio versus the redshift in Fig. 2. 
 There is no clear trend for 
 spectral evolution with redshift. 
We then examined whether the flattening of the spectrum is due to
spectral evolution with luminosity. We plot the hardness ratio versus
the 2-10 keV luminosity in Fig. 3. Again no trend for spectral
evolution with luminosity is apparent.

\begin{figure*}
\rotatebox{270}{\includegraphics[height=11.0cm]{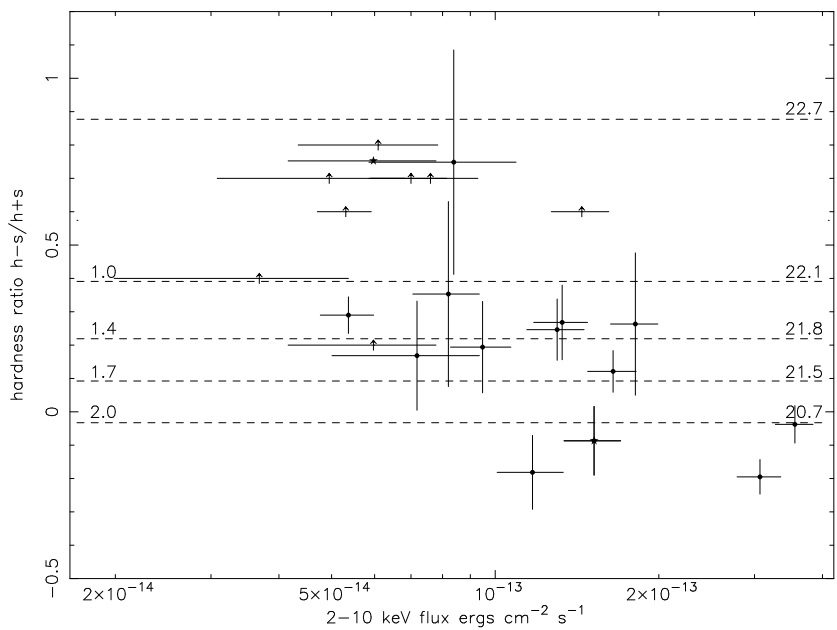}}
\caption{The hardness ratio (HR) versus flux for our QSOs with 
predicted HRs for various power-law indices(left hand scale) and for a
power-law with $\Gamma=1.9$ and a range of absorption (right hand
scale). The two with possible radio counterparts are shown (stars). 
Arrows indicate lower limits.}
\end{figure*}

\subsection{The integrated QSO spectrum}

Here we derive the average X-ray spectrum 
stacking together the QSOs photons in each field.
Firstly we fit the \asca data alone in the 0.8-8.0 keV energy band, 
forcing the two GIS detectors to
have the same normalisations and we tie the spectral index for the power-law
component to take the same value in all data sets.
We find that a single power law (PL) with $\Gamma=1.56\pm0.18$
for $\chi^2=104.89/99$ degrees of freedom (d.o.f.) with  
the hydrogen column density fixed to the Galactic value (in the range
of $1.7\times10^{20} \rm cm^{-2}-1.9\times10^{20} \rm cm^{-2}$ ) is 
a reasonable fit, in perfect with the hardness ratio
analysis.

We then fit the \rosat data.
Again, we 
tie the spectral index for the power-law
component to take the same value in all data sets, over the 0.5-2.0 keV range.
We fit the data with a single power-law component and we obtain a
slope of 
$\Gamma =2.32 \pm 0.2$ ($\chi^2 = 170.23/173$ d.o.f.). 
Over the full 0.1-2.0 keV range a single power-law gives a poor
fit ($\Gamma = 2.44$ with $\chi^2 = 690/333$ d.o.f.).

A joint \asca-\rosat fit over the 0.5-8.0 keV energy range was
carried out as well. The single power-law fit gives 
$\Gamma=2.16\pm0.1$ 
($N_H=$ Galactic and $\chi^2=309.21/273$ d.o.f.). However, the  obtained
spectral index is
clearly discrepant with the \asca data alone, while it is in good
agreement with the \rosat data. This is attributed to
the better statistics of the \rosat data compared to  \asca, which
drive the spectral fitting, while it has been suggested that the
spectral index discrepancy may be due to calibration
 uncertainties  in the PSPC and GIS response matrices 
causing a fake soft excess (Iwasawa, Nandra, Fabian 1999).
 We have parameterised this spectral upturn with a broken
power-law model (BKN PL model). In this case the soft and hard 
 components of the
spectrum are represented by two power-law with spectral indices
$\Gamma_1$ and $\Gamma_2$, respectively. We obtained a good fit with $\Gamma_1=2.35^{+0.11}_{-0.11}$
and $\Gamma_2=1.47^{+0.22}_{-0.20}$ and a break point for the energy of 
$1.48^{+0.32}_{-0.19}$ keV. It is clear that these are in
agreement with the indices obtained by fitting the data of each
detector alone. 

\begin{figure}
\rotatebox{270}{\includegraphics[height=8.0cm]{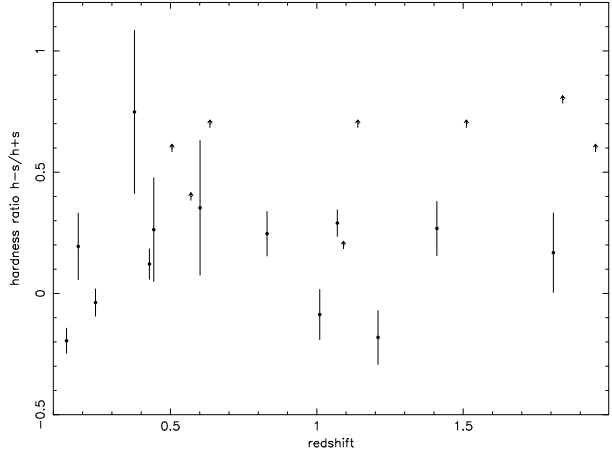}}
\caption{The HR versus the redshift of the sources. }
\end{figure}

\begin{figure}
\rotatebox{270}{\includegraphics[height=8.0cm]{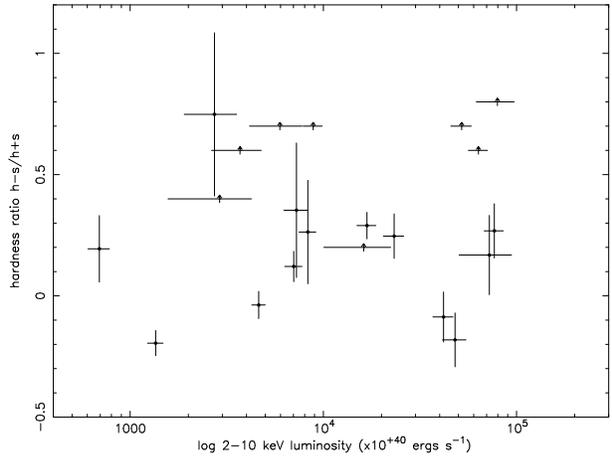}}
\caption{The HR versus the 2-10 keV luminosity. }

\end{figure}

We then added 
a soft-excess black-body component with kT=0.1 keV to the joint
\rosat-\asca fit (PL+BB model). In this case we obtain a good fit as well
($\chi^2=271.46/261$ d.o.f.), with $\Gamma=1.71^{+0.19}_{-0.17}$. The
value of the slope is consistent within the 90 percent errors bars,
with the slope obtained by fitting the \asca data alone.
The parameters derived when including the blackbody component are
in excellent agreement with the findings of Blair \etal (2000). The
above authors derived the average ROSAT spectrum of soft X-ray 
selected QSOs using 150 objects.
They splitted the sample in five redshift bins and 
they obtained a power-law slope of
$\Gamma=1.8-1.9$ (and a soft excess with kT=0.1 keV), in all  bins. 
We emphasize that although the above models parameterise
successfully the data, no real physical
significance should be attributed. 
This is due to the fact that our data span a wide range of
redshifts
and therefore the resulting spectrum is a blend
 of various spectral features at different rest-frame energies.  
However, the requirement of the data for this component clearly
demonstrates the need for a spectral upturn at low
energies (below $\sim$ 2 keV).  
A summary of the fits is shown in Table 3.

\begin{figure*}
\begin{center}
\begin{tabular}{ll}
{\large {(a)}} &  {\large {(b)}} \\
\rotatebox{270}{\includegraphics[height=8.0cm]{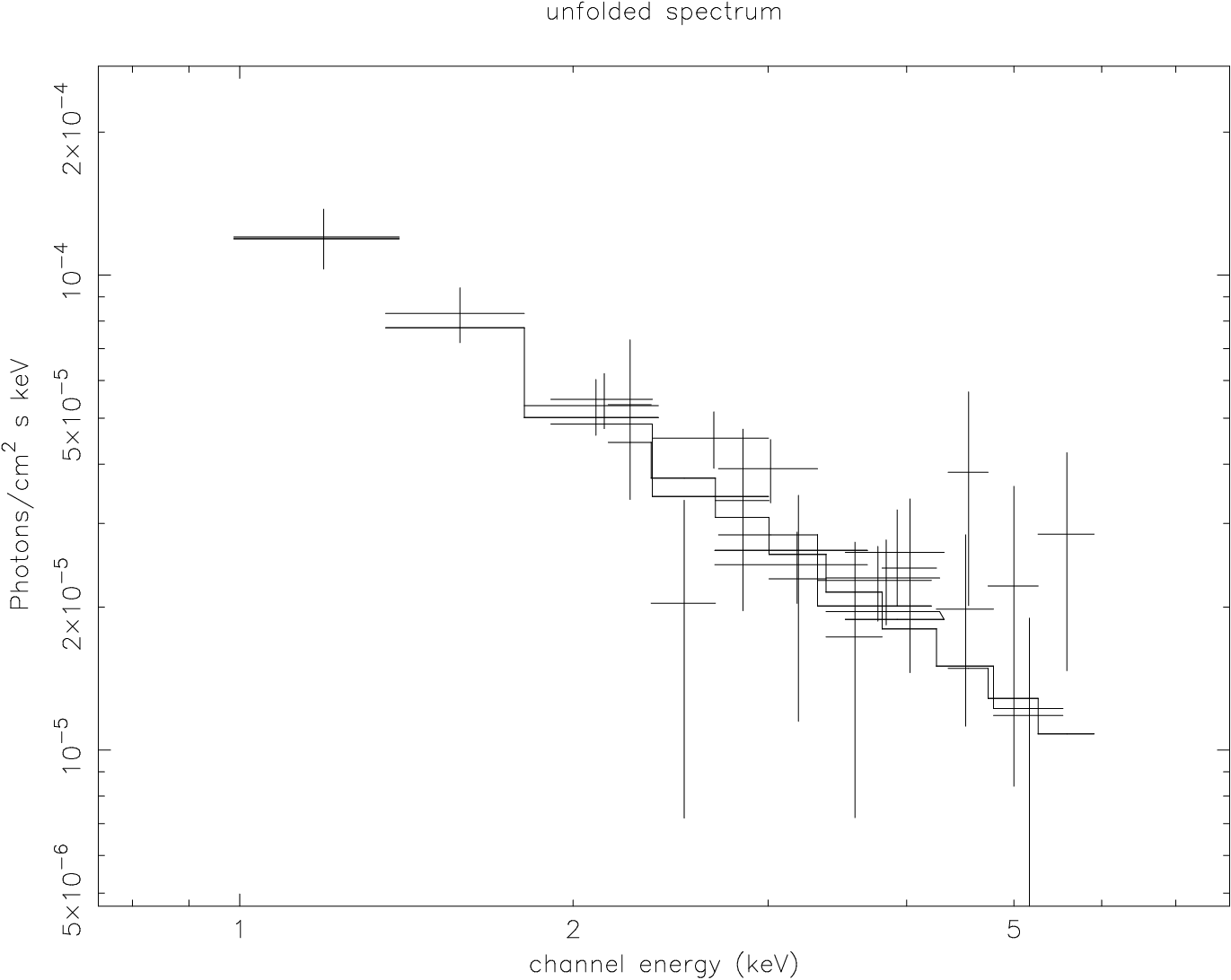}}
 &\rotatebox{270}{\includegraphics[height=8.0cm]{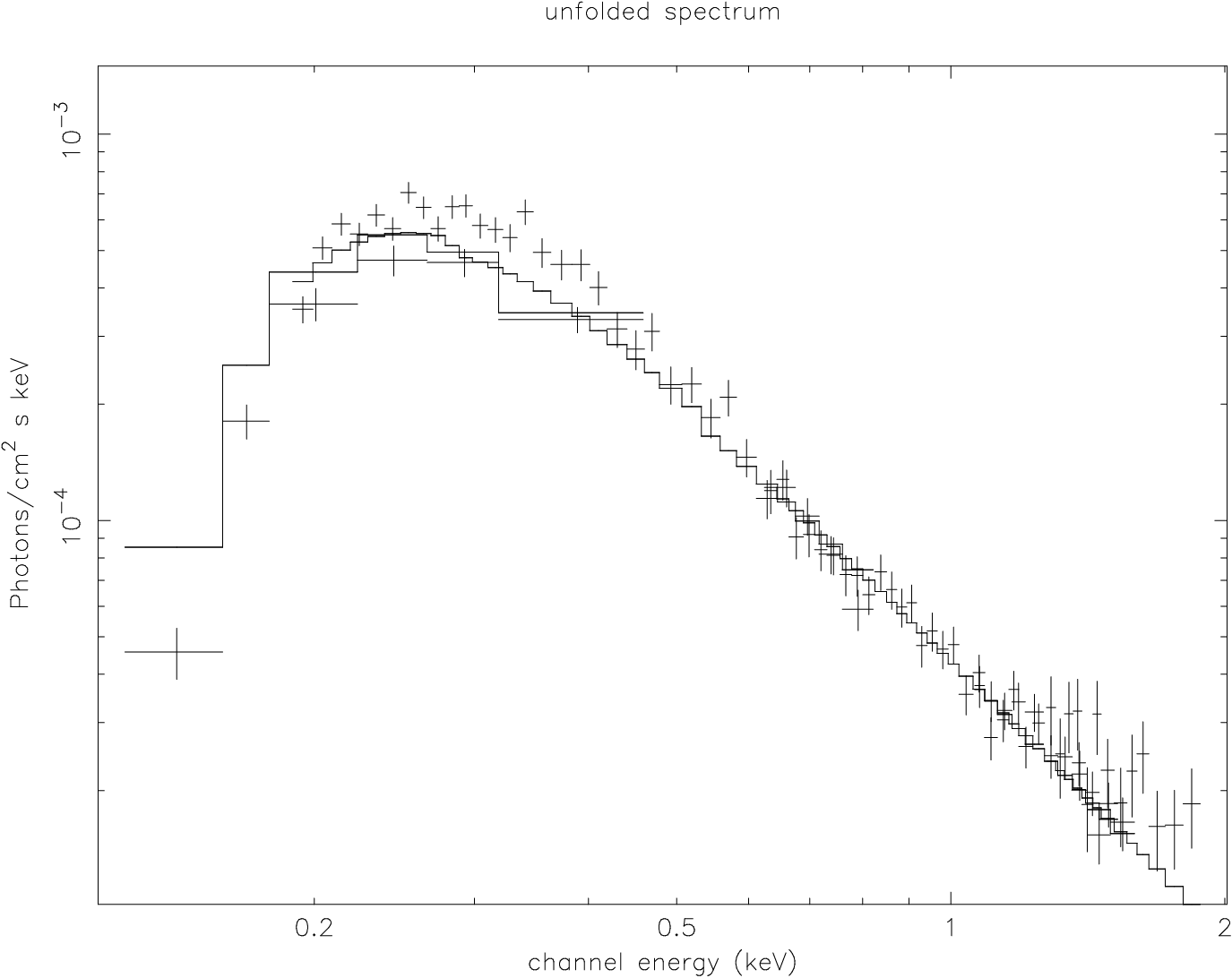}}\\

{\large {(c)}} & {\large {(d)}}  \\
\rotatebox{270}{\includegraphics[height=8.0cm]{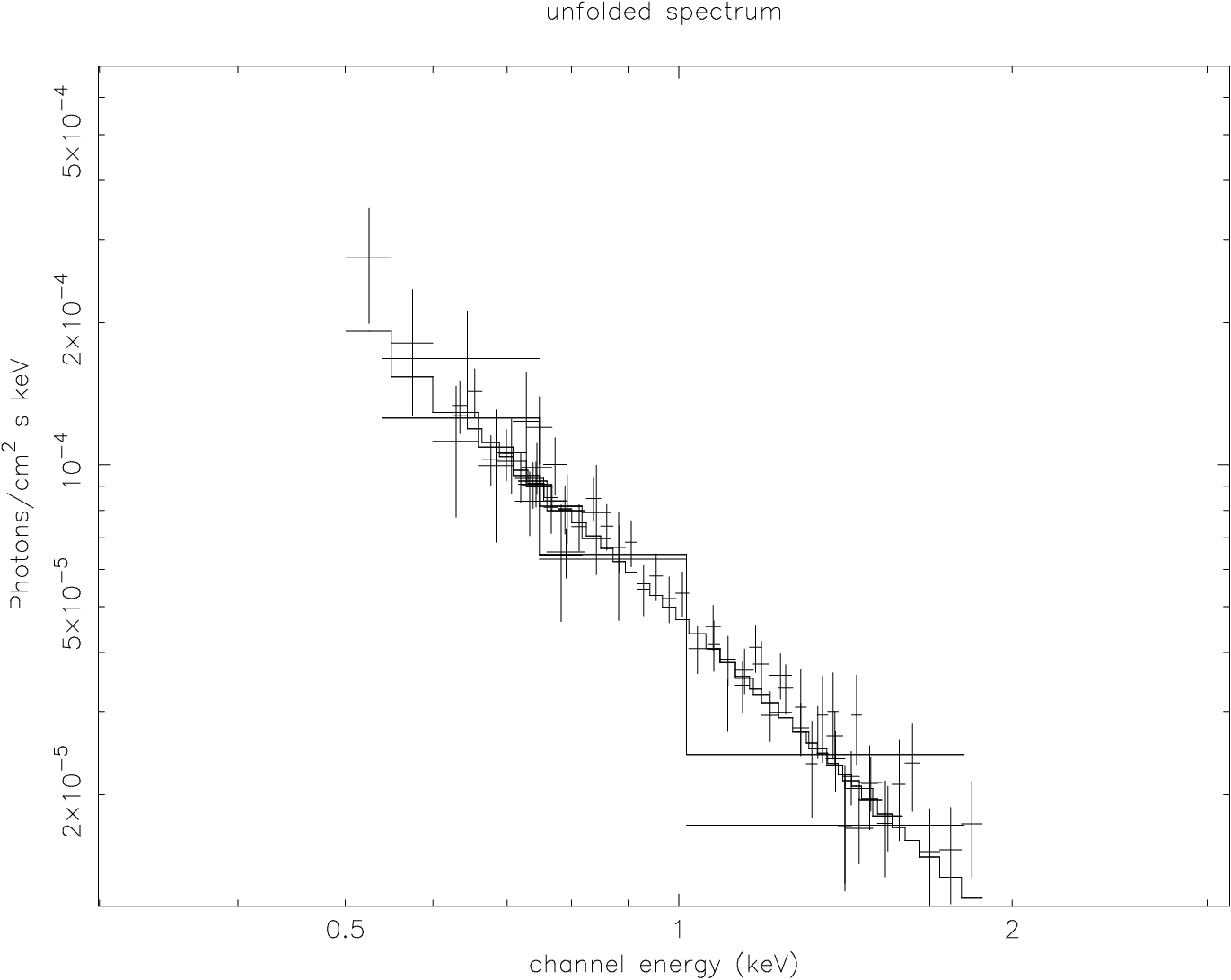}}
&
\rotatebox{270}{\includegraphics[height=8.0cm]{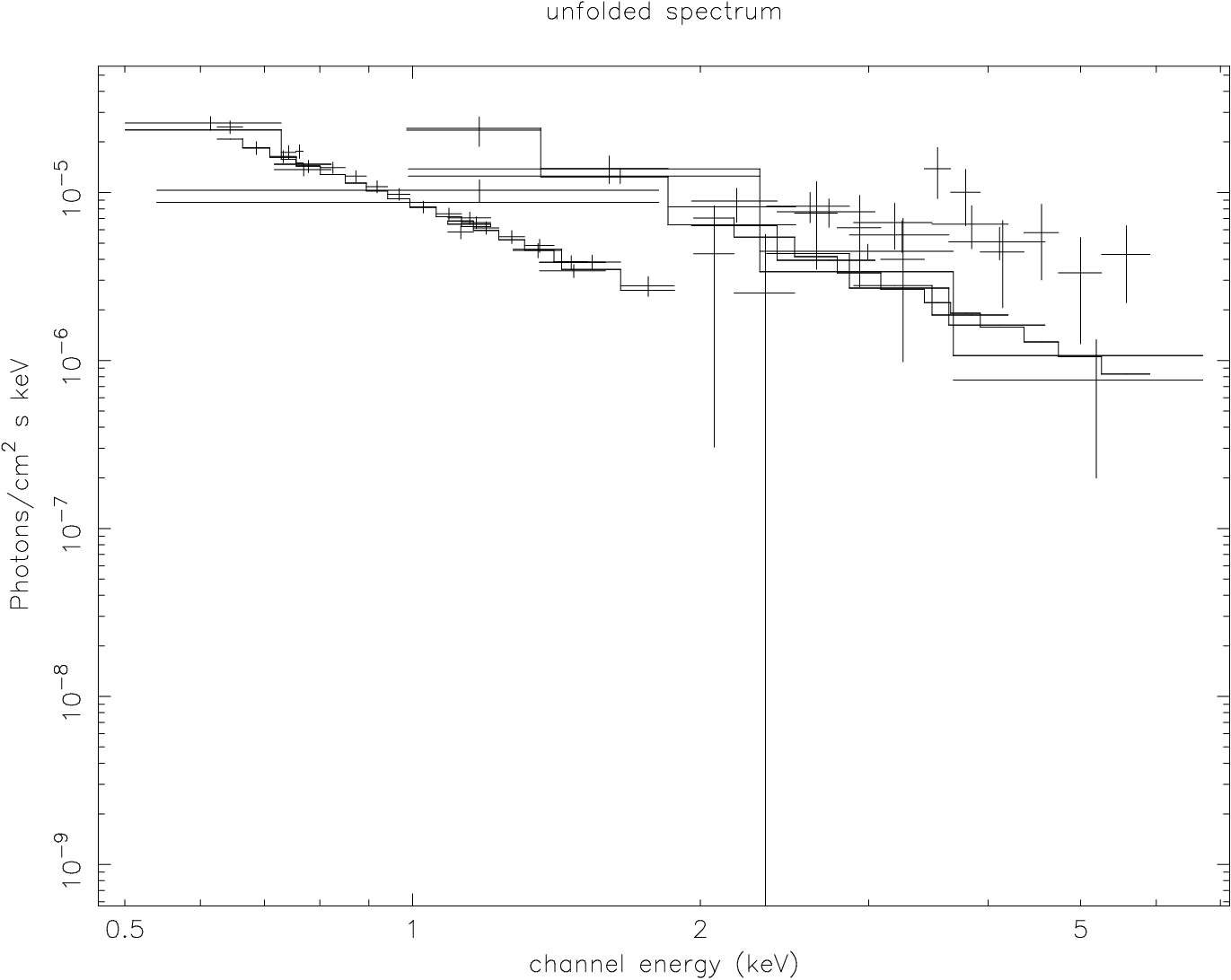}} \\

\end{tabular}
\end{center}

\caption{The unfolded spectrum for the power-law model for (a) the \asca
data over the 0.8-8.0 keV; (b) the \rosat data over the 0.1-2.0 keV; (c)\rosat data over the
0.5-2.0 keV and (d) the joint data over the 0.5-8.0 keV. The model fits are shown
by solid histogram.}  
\end{figure*}

\begin{table*}
\caption{Results of spectral fits, fixing the photoelectric absorption
to the galactic values shown in Table 1. } 
\begin{tabular}{ccccccc}  

           & \asca & \rosat & \rosat &\asca-\rosat& \asca-\rosat & \asca-\rosat \\
           &   PL  &  $PL^{a}$    &    $PL^{b}$       & PL   &  PL+BB       &  BKN PL       \\\hline
$\Gamma$   &$1.56\pm0.18$ &$2.32\pm0.22$&$2.44^{+0.01}_{-0.02}$ &$2.16\pm0.1$&$1.71^{+0.19}_{-0.17}$&$2.35\pm0.11$\\
kT or E(keV)&- &- &-& -&0.1&$1.48^{+0.32}_{-0.19}$\\
$\Gamma_2$ &-&-&-&-&-&$1.47^{+0.22}_{-0.20}$\\
$\chi^2(dof)$&104.89(99)&170.23(173)& 690.69(333)&309.21(273)&271.46(261)&276(271)\\\hline

\end{tabular}

$^{a}$spectral fitting in the 0.5-2.0 keV band \\
$^{b}$spectral fitting in the 0.1-2.0 keV band
\end{table*}



\section{Discussion and Conclusions}

 We investigate the properties of the typical, high redshift, 
  hard X-ray selected QSOs using 
 a sample of 21 objects detected in six deep 
 \asca fields.
 The hardness ratio analysis showed that the source spectra harden
 towards fainter fluxes. 
The majority of the objects have indices $\Gamma\leq$1.7. 
 This flattening has been previously reported by 
 Ueda \etal (1999) and  Della Ceca \etal (1999).
 However, unlike the above,  our sample consists exclusively of QSOs.
Therefore it appears that over the hard band the QSOs show evidence for
spectral hardening towards fainter fluxes.

The spectral analysis of the stacked spectrum    
shows that the \asca yields an effective  spectral index of
$\Gamma=1.56\pm0.18$, for type-1 QSOs. 
This is lower than 
the canonical spectral index of
bright, nearby  QSOs (eg Reeves et al. 1997, Lawson et al. 1998). 
In contrast, the above spectral index is more consistent with 
the XRB spectrum (Gendreau 1995). Page  (1997) studied the stacked spectrum 
 of 34 {\it soft X-ray selected} QSOs from the RIXOS \rosat survey using 
\asca.  
 He finds a spectral index of $\Gamma=1.8\pm 0.1$ consistent with 
 that of QSOs in the 2-10 keV band. 
 At first  glance, the spectral flattening of the hard X-ray 
 selected QSOs witnessed here 
 could be due to intrinsic absorption. Indeed, 
 a few high redshift QSOs with large amounts of absorption 
 have been found so far in hard X-ray surveys (eg Fiore et al. 1999).   
 However, in our case the \rosat data do not favour this possibility. 

 The stacked \rosat spectrum alone has a spectral index $\Gamma \approx2.3$ 
 consistent with the spectral index of individual soft X-ray selected 
 QSOs in the soft 0.1-2 keV band (eg Laor et al., 1997, Fiore et al. 1998). 
 Consequently, when we fit simultaneously the \rosat 
 and \asca spectra, the data 
 suggest the presence of a spectral upturn at soft energies. 
 This upturn can be naively modeled with a black body component 
 with $\rm kT\sim 0.1$ keV, although  this  
 obviously has no physical interpretation. 
 Then the power-law index becomes $\Gamma=1.7\pm 0.2$, in good 
 agreement with the spectrum of nearby QSOs. 
 Here, we note that Iwasawa, Nandra \& Fabian (1999) raised some 
 questions about the consistency of the \rosat and 
 \asca spectral fits. In particular they fitted  
 simultaneous  \rosat and \asca spectra of NGC5548
 finding that the \rosat spectrum can be steeper by as 
 much as $\Delta\Gamma\sim 0.4$ 
 in the common 0.5-2 keV band. This discrepancy could be possibly attributed 
 to uncertainties in the calibration of both \rosat and \asca instruments.  
 Another  possibility is that the concave spectrum obtained 
 is an artefact of our assumption 
 that the spectra of QSOs can be represented by a single value of $\Gamma$.
 More specifically the co-addition of spectra with steep and flat spectral indices 
 may result 
 in a steep and flat spectrum at low and high energies respectively. 
 A final possibility is that the QSO spectra are indeed concave as
originally suggested by Schwartz $\&$ Tucker 1988 (see also Mineo
\etal 2000).

However, it is puzzling that in contrast with our results as well as the other \asca and \bepposax  results, \chandra has not found a large number of broad line QSOs with flat X-ray 
spectrum. In their hard selected sample only 2 QSOs were found. 
These have a spectral index $\Gamma$  of  $\sim$1.8. 
(Mushotzky \etal 2000). Possibly this is due to the sensitivity curve 
of \chandra which peaks below 2 keV and is lower than the \asca
sensitivity above $\sim$ 7 keV and, therefore, possibly biases the 
results in favour of steep spectrum X-ray sources in the manner suggested
by Della Ceca \etal (1999).

  \xmm with its high effective area as well its broad bandpass 
 (0.1-12 keV) is expected 
 to  shed more light on the
 spectrum of individual, faint, high redshift QSOs 
 which produce the largest fraction of the XRB. 

\section{Acknowledgments}
This research has made use of data obtained from the Leicester Database and 
Archive Service at the Department of Physics and Astronomy, Leicester 
University, UK.

\end{document}